\documentclass[aps,prl,preprint]{revtex4}

\usepackage{graphicx} 
\usepackage{amsmath}
\usepackage{stmaryrd}
\usepackage{MnSymbol}
\usepackage{setspace}
\usepackage{physics}
\usepackage{mathtools}
\usepackage{amsfonts}
\usepackage{color}
\usepackage{cancel}

\newcommand{\qp}[1]{\hat{#1}}
\newcommand{\bv}[1]{\boldsymbol{#1}}

\newcommand{\vc}[1]{\textbf{#1}}

\begin{document}

\title{A numerical study of the spatial coherence of light in collective spontaneous emission}
\author{D. D. Yavuz, A. Yadav, D. C. Gold, T. G. Walker, and M. Saffman}
\affiliation{Department of Physics, 1150 University Avenue,
University of Wisconsin at Madison, Madison, WI, 53706}
\date{\today}
\begin{abstract}

We present a numerical study of the spatial coherence of light that is radiated from a dilute ensemble of atoms. The spatial coherence is established as a result of the collective (cooperative) coupling of the atoms to the light, and is qualitatively different from the coherence of a laser. Specifically, the coherence in collective spontaneous emission does not rely on population inversion and stimulated emission, is governed by anti-phasing of the dipoles (subradiance), and the key figure-of-merit for the observed coherence is $\sim\frac{N}{(L/\lambda)}$. Here, $N$ is the number of atoms in the ensemble, $L$ is the size of the sample, and $\lambda$ is the wavelength of the emitted light.

\end{abstract} 
\maketitle

It has been well-known since the pioneering work of Dicke \cite{dicke} that the physical process of spontaneous emission can be importantly modified as a result of the coherent interaction of the radiators \cite{haroche,eberly,scully1,yelin,francis,adams,jenkins,ritsch,bachelard,petrov,zanthier,agarwal,reitz}. For example, when we have $N$ atoms tightly packed in a small volume that are excited to a radiating state,  their emissions can spontaneously synchronize, producing a fast decay rate. This is typically referred to as Dicke superradiance, and has been observed in a large number of physical systems including neutral atoms \cite{bloch,an,gauthier,kuga}, ions  \cite{ions}, molecules \cite{molecules}, nitrogen vacancy centers \cite{diamond1,diamond2}, and superconductors \cite{superconducting}. Superradiance relies on constructive (in-phase) interference of the emissions. Under certain conditions, the emitted light from individual radiators can also interfere destructively (out-of-phase interference), producing a reduced decay rate, which is referred to  as subradiance.

While most early studies of collective spontaneous emission focused on dense ensembles in the Dicke limit (with a large number of atoms per cubic wavelength of volume) \cite{feld,manassah},  much recent work has investigated subradiance near or outside the Dicke limit \cite{kaiser1,kaiser2,kaiser3,kaiser4,kaiser5,browaeys1,browaeys2,browaeys3}. Recent studies have shown that even in the dilute regime (i.e., very few number of atoms per cubic wavelength of volume), collective coupling of the emitters can play an important role. For example, we have recently demonstrated that collective decay is even relevant in dilute clouds with a very low optical depth \cite{dipto,davidexp}. Specifically, we have experimentally observed that the decay rates can be reduced due to subradiance by as much as 20~\%, inside a dilute cloud at an optical depth of $10^{-2}$ or less. In this regime, subradiant states that are correlated across the whole ensemble are the dominant decay mechanism, as evidenced by the spatial coherence of the emitted light. By coupling the emitted light to a misaligned Michelson interferometer, we studied the spatial coherence of the emission and investigated the dependence of the spatial coherence on the number of atoms as well as the atomic temperature \cite{davidexp}. 

In our recent experimental paper, we provided a qualitative explanation for the observed results \cite{davidexp}. In this work, we present a numerical study of the spatial coherence of light in collective spontaneous emission, in the same regime of dilute ensembles with a very low optical depth. As we discuss in detail below, we consider a disordered ensemble in three dimensions with random atomic positions. The atomic ensemble is initially excited with a weak resonant laser beam. After the laser beam is switched off, we first calculate the evolution of the collective dynamics using the dipole-dipole exchange Hamiltonian. As the atoms are undergoing collective spontaneous emission, we then calculate the spatial profile of the  emitted fluorescent light in the far-field, along a direction orthogonal to the laser propagation direction. Initially, the light emitted from individual atoms along the direction orthogonal to the laser propagation is uncorrelated due to random locations of the atoms in the disordered ensemble. As a result, the initial spatial pattern of the emitted light resembles a random speckle pattern, displaying little spatial coherence. As the system evolves and dipole-dipole correlations build up, the collective wavefunction is driven into subradiant states that have long range correlations across the ensemble. Emission from these subradiant modes then displays long-range correlations; i.e., light with spatial coherence across length scales comparable to the size of the ensemble. We  investigate how the spatial coherence evolves as a function of time, and also study the dependence of the coherence as the parameters of the atomic ensemble, such as the number of atoms is varied. 

As we will detail below, an important result of our numerical study is that we confirm the relevant figure-of-merit for collective spontaneous emission in dilute ensembles with a very low optical depth. This figure-of-merit is not the number of atoms, or the atomic density, or the optical depth of the atomic ensemble. Instead, for a symmetrical 3D sample, the relevant figure-of-merit is $\sim \frac{N}{(L/\lambda)}$ ($N$ is the number of atoms in the ensemble, $L$ is the size of the sample, and $\lambda$ is the wavelength of the emitted light). This figure-of-merit was qualitatively discussed in our recent experimental work \cite{davidexp}. It is also implicit in the recently derived width of the eigenvalue distribution of the exchange Hamiltonian \cite{ben}. This figure-of-merit differentiates collective decay in dilute ensembles from subradiance in optically-thick clouds \cite{kaiser1,kaiser2,kaiser3,kaiser4} and the traditional understanding of large-sample superradiance \cite{haroche,eberly}. For sub- and super-radiance, the important parameter is the optical depth of the ensemble, which would be $\sim \frac{N}{(L/\lambda)^2}$ for a symmetrical 3D sample. 

The spatial coherence of light that is produced during collective spontaneous emission in dilute atomic clouds is distinctly different from the spatial coherence of a laser. Specifically, there are three key differences \cite{siegman}: (1) The spatial coherence established in collective decay does not rely on population inversion which is critically required for the operation of a laser. (2) In lasers, the phasing of the atomic dipoles through stimulated emission is necessary. In contrast, in dilute ensembles experiencing collective decay, subradiance (out-of-phase superpositions) is the dominant mechanism that establishes correlations between the atomic dipoles. (3) In stimulated emission, the gain-length product of the medium (which is proportional to the optical depth) is the key figure-of-merit. In contrast, as we mentioned in the previous paragraph, in spatial coherence due to collective spontaneous emission, the relevant figure-of-merit is $\sim \frac{N}{(L/\lambda)}$. 

In recent years, there has been an increased interest in collective spontaneous emission, in particular within the context of quantum information science.  Some recent highlights of theory work include highly-directional mapping of quantum information between atoms and light in two-dimensional arrays \cite{yelin1,yelin2}, studies of broadening and photon-induced atom recoil in collective emission \cite{francis1,francis2,francis3}, light storage in optical lattices \cite{ritsch2,ballantine,garcia}, collective nonclassical light emission and hyperradiance \cite{agarwal2,agarwal3,agarwal4}, and improving photon storage fidelities using subradiance \cite{kimble}. On the experimental front, as we mentioned above, much early work on subradiance used disordered ultracold atomic clouds \cite{kaiser1,kaiser2,kaiser3,kaiser4,kaiser5,browaeys1,browaeys2,browaeys3}, including our recent work which used dilute ensembles with low optical depth \cite{dipto,davidexp}. Recent experiments using ultracold atoms have demonstrated single atomic layer mirrors \cite{bloch}, phase transitions \cite{browaeys4}, as well as enhanced collective coupling using optical cavities \cite{yan}.

\section{Dipole-dipole interaction and the exchange Hamiltonian}

When an atomic ensemble undergoes collective decay, one approach to model the dipole correlations that build up across the sample is through the reduced atomic-only exchange Hamiltonian. This Hamiltonian is obtained by tracing out the radiation coordinates, and has been used by other authors to study the physics of collective decay \cite{kimble,francis}. A detailed derivation of this Hamiltonian has been discussed, for example, in Ref. \cite{ben}. Briefly, the full Hamiltonian of the whole system consisting of $N$ two-level atoms interacting with a continuum of radiation modes is given by ($\hbar=1$) \cite{haroche,eberly},
\begin{eqnarray}
\hat{H}=\sum_{j=1}^{N}\frac{\omega_{a}\hat{\sigma}^{j}_{z}}{2}+\sum_{\vc{k},\bv{\epsilon}}\omega_{\vc{k},
\bv{\epsilon}}\left(\hat{a}^{\dagger}_{\vc{k},\bv{\epsilon}}\hat{a}_{\vc{k},\bv{\epsilon}}+\frac{1}{2}\right)
-\sum_{j=1}^{N}\sum_{\vc{k},\bv{\epsilon}}(g^{*}_{\vc{k},\bv{\epsilon}}\,e^{i \vc{k}\cdot\vc{r}_{j}}\hat{\sigma}^{j}_{+}\hat{a}_{\vc{k},\bv{\epsilon}}+g_{\vc{k},\bv{\epsilon}}\,e^{-i \vc{k}\cdot\vc{r}_{j}}\hat{\sigma}^{j}_{-}\hat{a}^{\dagger}_{\vc{k},\bv{\epsilon}})
\end{eqnarray}
	where, 
\begin{eqnarray}
\hat{\sigma}_z^j & =& |1\rangle^j \hspace{0.1cm} {^j}\langle 1|-|0\rangle^j \hspace{0.1cm} {^j}\langle 0| \quad , \nonumber \\
\hat{\sigma}_+^j & =& |1\rangle^j \hspace{0.1cm} {^j}\langle 0| \quad , \nonumber \\
\hat{\sigma}_{-}^j & =& |0\rangle^j \hspace{0.1cm} {^j}\langle 1| \quad ,
\label{m2}
\end{eqnarray}

\noindent are the atomic spin operators for the $j$'th atom with energy eigenstates $|0 \rangle^j$ and $|1 \rangle^j$, respectively. The quantity $\omega_a$ is the atomic transition frequency. The operators $\qp{a}_{\vc{k},\bv{\epsilon}}$ and $\qp{a}^{\dagger}_{\vc{k},\bv{\epsilon}}$, are the photon annihilation and creation operators for the radiation mode with wave-vector $\vc{k}$ and polarization $\bv{\epsilon}$. The well-known Dicke limit can be obtained from the above Hamiltonian when the size of the sample is small compared to the radiation wavelength set by the relevant $k$, i.e., $\vc{k}\cdot\vc{r}_{j}\rightarrow 0,\forall\,\,0\le j\le N$ ($\vc{r}_{j}$ is the position of the $j$th atom). Using the Born-Markov approximation and tracing out the radiation modes, we can reduce the above full-Hamiltonian to the following atomic-only effective Hamiltonian $\qp{H}_{eff}$ \cite{ben}:
\begin{eqnarray}
\hat{H}_{eff}=\left(\omega_a-i\frac{\Gamma}{2}\right)\sum_{i=1}^{N}\frac{\hat{\sigma}^{i}_{z}}{2}+\overbrace{\sum_j \sum_{k\neq j} \hat{H}^{jk}}^{\qp{H}_{exc}} \quad .
\label{m3}
\end{eqnarray}

\noindent Here,  the dipole-dipole interactions between different atoms can be grouped as an exchange Hamiltonian, $\qp{H}_{exc}$. This Hamiltonian is a sum is over all pairs of atoms, and operators $\hat{H}^{jk}$ act nontrivially only on atoms with indices $j$ and $k\neq j$:
\begin{eqnarray}
\hat{H}^{jk}= F_{jk}  \hat{\sigma}_+^j  \hat{\sigma}_-^{k} +F_{kj}  \hat{\sigma}_-^{j}  \hat{\sigma}_+^{k} \quad .
\label{m4}
\end{eqnarray} 

\noindent The pairwise Hamiltonian $\hat{H}^{jk}$ is essentially a ``spin" exchange interaction (dipole-dipole interactions in the present case for a pair of atoms) which is mediated by photon modes with symmetric (but non-unitary) coupling constants of $F_{jk}$:
\begin{eqnarray}
F_{jk}= F_{kj} &=& -\left(i \frac{\Gamma}{2}\right) \left( \frac{3}{8 \pi} \right) \left[ 4 \pi  (1-\cos^2\theta_{jk}) \frac{\sin k_a r_{jk}}{k_a r_{jk}} 
+ 4 \pi (1-3\cos^2\theta_{jk})  \left(\frac{\cos k_a r_{jk}}{(k_a r_{jk})^2} -\frac{\sin k_a r_{jk}}{(k_a r_{jk})^3} \right)   \right] \quad . 
\label{m5}
\end{eqnarray}

\noindent Here, $\Gamma$ is the decay rate of a single isolated atom. $r_{jk}$ is the distance between the two atoms, and $\theta_{jk}$ is the angle between the atomic dipole moment vector and the separation vector $\vc{r}_{jk}$. The quantity $k_a$ is the wave vector for the radiation modes that are energy-resonant with the atomic transition: $k_a = \omega_a /c $.

\section{General set-up of the simulations}

The simplified schematic of the numerical simulations is shown in Fig.~1. We consider a disordered three dimensional ensemble where the atoms are distributed randomly in a certain volume, with dimensions $L_x$, $L_y$, and $L_z$, respectively. The ensemble is excited with a laser beam propagating along the $x$ direction and the emitted light is collected at an axis orthogonal to the laser propagation: the $z$ axis. The laser can be polarized along the $y$ or the $z$ axis. For concreteness, we will take the laser to be polarized along the $y$ axis, which also sets the direction of the dipoles. We have verified that the results are very similar if the laser is instead polarized along the $z$ axis. 

\begin{figure}[tbh]
\vspace{0cm}
\begin{center}
\includegraphics[width=0.9\textwidth]{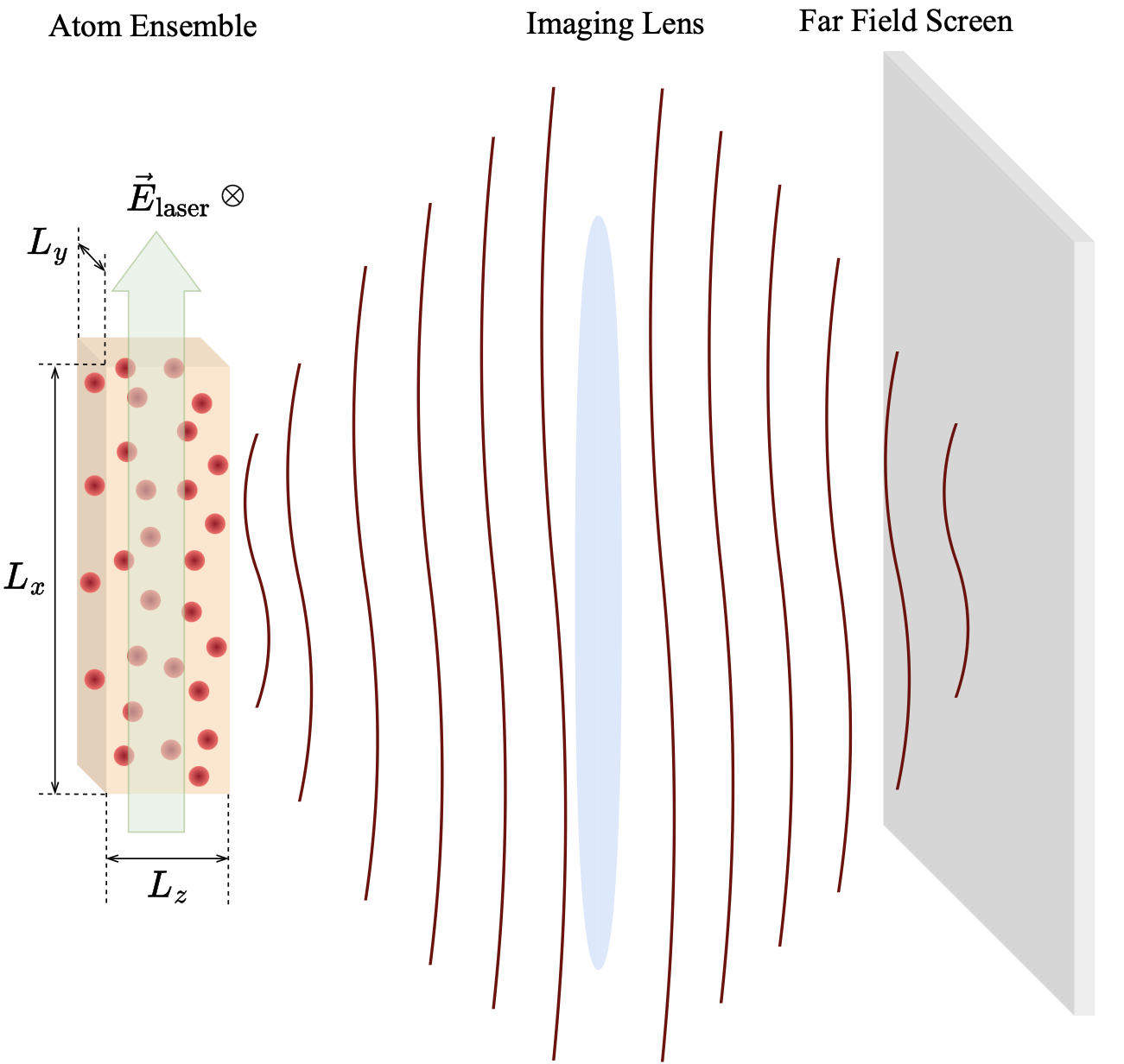}
\vspace{-0.2cm} 
\begin{singlespace}
\caption{\label{scheme} \small  The simplified schematic of the numerical simulations.  We consider a three-dimensional atomic ensemble where the atoms are distributed randomly in a certain volume. The ensemble is excited with a weak laser beam propagating along the $x$ direction and the emitted light is collected at an axis orthogonal to the laser propagation direction (the $z$ axis). The laser excites the atoms and is then abruptly turned off. As the ensemble undergoes collective decay and the correlations build-up between the dipoles, we calculate the full three dimensional dipole emission from each radiating atom in the far-field. This far-field is then imaged using a lens to the imaging plane. We take imaging to be ideal and free-of aberrations, and with a magnification of $M=1$.}
\end{singlespace}
\end{center}
\vspace{-0.cm}
\end{figure}

The laser beam excites the atoms and is then abruptly turned off. With the laser beam turned off, the atoms evolve under the effective Hamiltonian, $\hat{H}_{eff}$ which is given by Eq.~(3) of above. As the system is evolving and the correlations are building up between the dipoles, we calculate the full three dimensional dipole emission from each radiated atom in a manner that we will describe in detail below. The total far-field electric field is the sum of the emitted waves from each atom. This far-field radiated pattern is then imaged using a lens to the imaging plane. We take imaging to be ideal and free-of aberrations, and with a magnification of $M=1$; i.e., in the image plane we observe a one-to-one spatial image of the radiating atomic ensemble. 

We take the excitation laser beam to be weak and restrict the problem to the single excited subspace with the following basis of states:
\begin{eqnarray}
|q \rangle = |100...0 \rangle, \quad  |010...0 \rangle, \quad  |001...0 \rangle,  \quad ...  |000...1 \rangle. 
\end{eqnarray}

This assumption is critical as it reduces the size of the problem from the exponentially large Hilbert space dimension of $2^N$ to $N$. We will discuss how we expect the results to differ when one goes beyond the single-excited subspace, in the conclusions section below. With these definitions and approximations, we expand the total wavefunction of the atomic ensemble in the above-mentioned basis as:
\begin{eqnarray}
|\psi (t)\rangle =e^{-i (\omega_a-i\Gamma t/2)}\sum_{q=1}^{N}c_{q}(t)|q \rangle \quad ,
\label{a1} 
\end{eqnarray}

\noindent where the expansion coefficients, $c_q(t)$ evolve in time according the Schr\"odinger's equation:
\begin{eqnarray}
i  \frac{d |\psi(t) \rangle }{dt} = \hat{H}_{eff} |\psi(t) \rangle\implies i  \dot{\vc{c}}(t) = \hat{H}_{exc} \vc{c}(t) \text{ or, }\vc{c}(t)=\exp(-it\qp{H}_{exc})\;\vc{c}(0) \quad.
\label{a4}
\end{eqnarray}

\noindent Here, the quantity $\vc{c}(t)$ is a column vector of dimension $N$ containing the coefficients $c_q(t)$ as its' entries. 

We take the laser intensity to be uniform, with a size much larger than the size of the atomic cloud. As a result, each atom in the ensemble initially has the same probability of being in the excited state, but with the phase of the propagating laser imprinted on the initial excitation amplitude, $c_q(t=0)$:
\begin{eqnarray}
c_q(t=0) = \frac{1}{\sqrt{N}} \exp( i  k_{laser} x_q) \quad . 
\end{eqnarray}

\noindent Here, $x_q$ is the $x$ position of the excited atom in state $|q \rangle$. The quantity $k_{laser}$ is the wave-vector of the laser, $k_{laser} = \omega_{laser} / c = 2 \pi / \lambda_{laser}$. For simplicity, we will take the laser beam to be near-resonant with the two-level atomic transition and will therefore assume, $\omega_{laser} = \omega_a$ and  $k_{laser} = k_a$.  

\begin{figure}[tbh]
\vspace{-0cm}
\begin{center}
\includegraphics[width=1\textwidth]{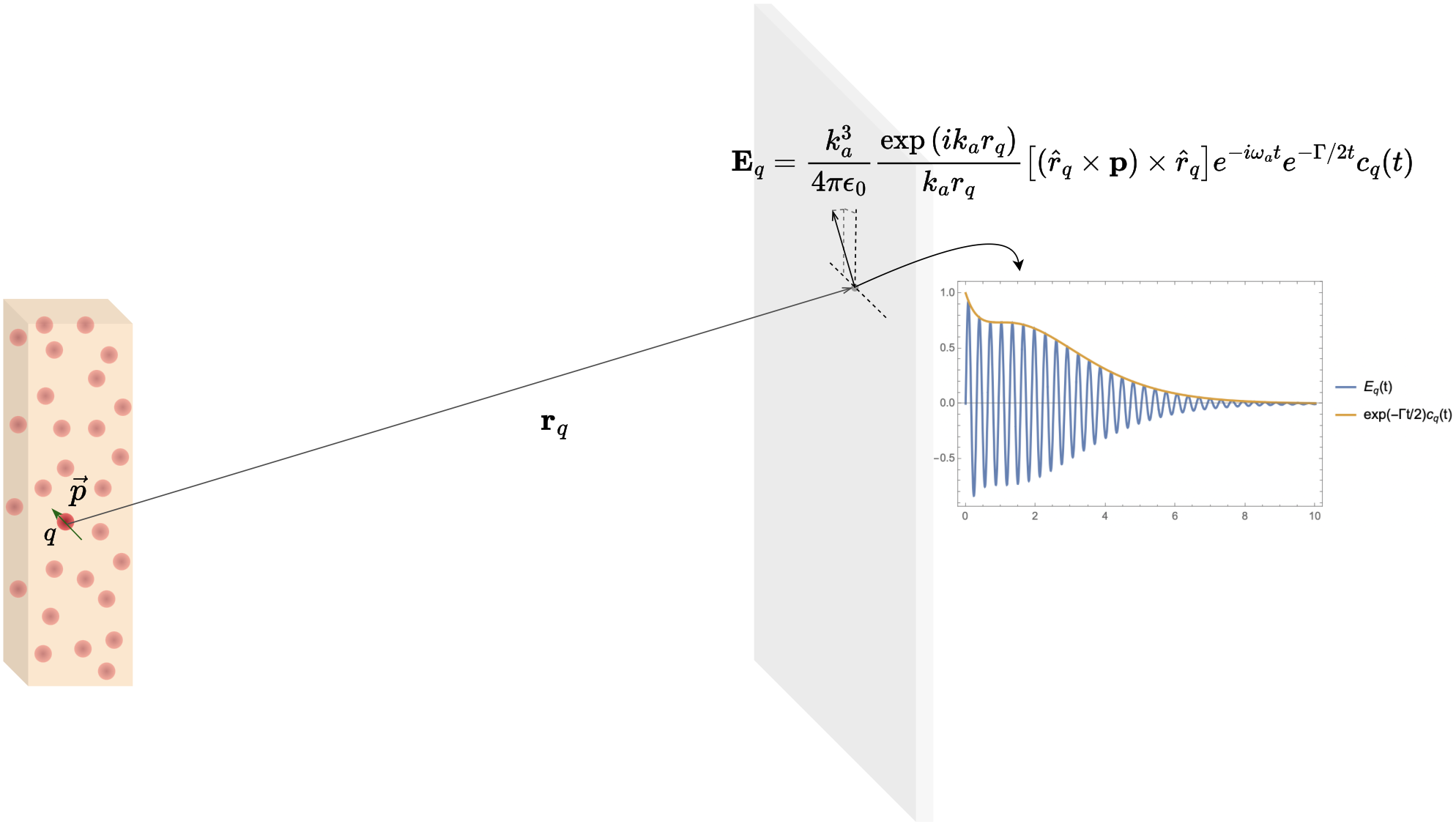}
\vspace{-0.2cm} 
\begin{singlespace}
\caption{\label{scheme} \small  The schematic for calculating the far-field radiation pattern from each dipole.  $\vc{p}$ is the pointing direction of the dipole, $r_q$ is the distance between the radiating atom in state $| q \rangle$ and that specific observation point, and the unit vector $\hat{r}_q$ is the direction that connects the radiating atom to the point on the observation plane. With the electric field of the emission from each atom calculated, the total field is found by using a coherent sum of all the emitted fields. }
\end{singlespace}
\end{center}
\vspace{-0.cm}
\end{figure}

We next discuss the calculation of the far-field radiation pattern, which is schematically displayed in Fig.~2. While the system is evolving, each atom can be thought of as a radiating dipole of magnitude $| \vc{p} | $ (which is given by the dipole matrix element of the two-level transition), with time varying amplitude, $c_q(t)$. Because the observation plane is in the far-field, we only need to consider the far-field of the radiating dipole pattern and we can ignore the near-field contributions. This far-field electric field that is radiated from state $|q \rangle$ in the observation plane is:

\begin{eqnarray}
\vc{E}_q (t, x, y, z=z_{obs}) =  \frac{k_a^3}{4 \pi \epsilon_0} \frac{\exp{i k_a r_q}}{k_a r_q}  \left[ (\hat{r}_q \times \vc{p}) \times \hat{r}_q  \right]  e^{-i\omega_a t}e^{-\Gamma t/2}c_q(t)  \quad . 
\end{eqnarray} 

Here, the dipole-moment vector $\vc{p}$ is also the polarization direction of the laser (i.e., the $y$ axis for the simulations that we discuss below). The quantity $r_q$ is the distance between the radiating atom in state $| q \rangle$ and that specific observation point, and the unit vector $\hat{r}_q$ is the direction that connects the radiating atom to the observation point. A detailed derivation of Eq.~(10) is given in Appendix~A below. With the electric field from each radiating atom calculated using Eq.~(10), the total field in the observation plane, (which we refer to as $\vc{E}_{obs}$) is a coherent sum of all of the emitted fields:
\begin{eqnarray}
\vc{E}_{obs} (t, x, y, z=z_{obs})  = \sum_{q=1}^N \vc{E}_q (t, x, y, z=z_{obs})   \quad . 
\end{eqnarray}

At a certain observation time, $t=t_{obs}$, the spatial coherence length is calculated by finding the width of the equal-time cross-correlation function \cite{mandel}. The cross-correlation function can be calculated in two different directions in the image plane; either along the $x$ direction or along the $y$ direction. We refer to these two functions as $H_x(x)$ and $H_y(y)$, respectively, and they are calculated using :
\begin{eqnarray}
H_x(x) & = &  \langle \int \vc{E}_{obs} (t=t_{obs}, x', y, z=z_{obs}) \cdot \vc{E}_{obs}^* (t=t_{obs}, x'-x, y, z=z_{obs}) d x' \rangle_y  \quad , \nonumber \\
H_y(y) & =  & \langle \int \vc{E}_{obs} (t=t_{obs}, x, y', z=z_{obs}) \cdot \vc{E}_{obs}^* (t=t_{obs}, x, y'-y, z=z_{obs}) d y' \rangle_x \quad . 
\end{eqnarray} 

Here, the operations $\langle \cdots \rangle_x$ and $\langle \cdots \rangle_y$ correspond to taking the spatial average along each direction, respectively. In the simulations that we describe below, we define the coherence length of the radiated spatial pattern, $w$, to be $1/e^2$ radius of the cross-correlation functions $H_x(x)$ and $H_y(y)$ . As expected, we numerically find the behavior to be almost identical in the two transverse axes, $x$ and $y$. Therefore, we will only display the results along one transverse axis, specifically along the $x$ axis. 

\section{Numerical Simulation results}

\subsection{Symmetric ensemble: subradiant time evolution}

We start with a symmetrical ensemble with parameters similar to our recent experiment \cite{davidexp}. We consider a disordered atomic cloud with an equal size of $40 \lambda$ along each direction: i.~e., $L_x= L_y = L_z = 40 \lambda$. We take $N=1500$ atoms which are uniformly and randomly distributed within this volume. The ensemble is dilute with an atomic density of $n=N/(L_x \times L_y \times L_z)=0.023/\lambda^3$. Furthermore, the optical depth of the ensemble is also very low: O. D.$= n \sigma L_x =n \frac{\lambda^2}{2 \pi} L_x = 0.15$. Here, the quantity $\sigma$ is the on-resonant absorption cross-section of the two-level transition.
\begin{figure}[h]
\vspace{0cm}
\begin{center}
\includegraphics[width=18cm]{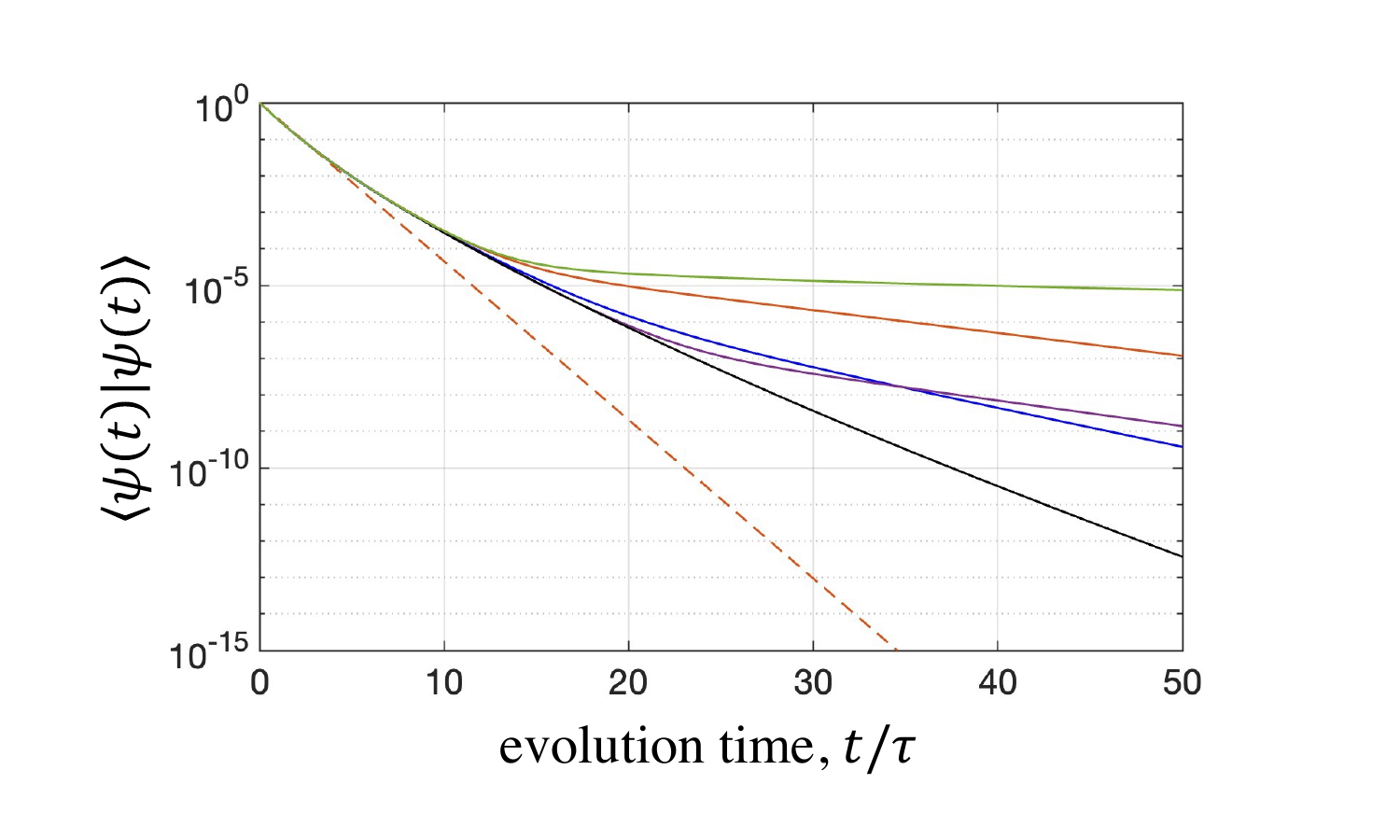}
\vspace{-1.2cm} 
\begin{singlespace}
\caption{\label{scheme} \small  Subradiant time evolution of a symmetric disordered ensemble with an atomic density of $n=N/(L_x \times L_y \times L_z)=0.023/\lambda^3$ and an optical depth of O.~D.$= n \sigma L_x =n \frac{\lambda^2}{2 \pi} L_x = 0.15$. The solid lines are the numerically calculated norm of the wavefunction, $\langle \psi (t) | \psi (t) \rangle$ (in logarithmic scale), while the dashed line is the exponential decay of an isolated individual atom, $\exp{(-t/\tau)}$. Initially, the decay curves are similar to each other and follow closely simple exponential decay. However as the dipole correlations build up, the collective decay becomes subradiant with ever slower decay rates at later times in the evolution. The difference that is observed in the time evolution is due to the difference in the initial randomly chosen locations of the atoms. }
\end{singlespace}
\end{center}
\vspace{-0.cm}
\end{figure}

Under these conditions, the collective decay is largely subradiant, with a time constant which itself evolves as a function of time. Figure~3 shows this result. Here we plot the norm of the wavefunction as a function of time for five different simulations, which are chosen to be representative examples. The solid lines are the numerically calculated $\langle \psi (t) | \psi (t) \rangle$ (in logarithmic scale), while the dashed line is the decay of an isolated individual atom, $\exp{(-t/\tau)}$ (the quantity $\tau$ is the lifetime of the excited state for an isolated individual atom, i.e., $\tau = 1 /\Gamma$). The difference that is observed in the time evolution (solid lines) is due to the difference in the initial randomly chosen locations of the atoms. Initially, the decay curves are similar to each other and follow closely to simple exponential decay (the dashed line). However as the dipole correlations build up, the collective decay becomes subradiant with ever slower decay rates at later times in the evolution. 

The large difference between the solid curves at later times in the evolution is due to the sensitivity of the dominant subradiant eigenstates of the system to the initial random distribution of the atomic positions. When one looks at the overall distribution of the eigenvalues of the Hamiltonian (i.e., the collective decay rates), the distribution will look similar for different random ensembles: centered around the independent decay rate $\Gamma$, with some eigenvalues lower than $\Gamma$ (subradiant) and some higher than $\Gamma$ (superradiant) \cite{ben,kimble}. While the distributions will look similar (as expected since we have a large number of atoms), the extreme eigenvalues (corresponding to most subradiant and most superradiant collective states) can differ substantially. This results in a large difference in the evolution of the wavefunction at later times, since the long-time-scale evolution is determined by the most subradiant modes. Not only the decay rates, but also the spatial structure of the extreme eigenstates are substantially different. This is the main reason for the large difference in the evolution of the spatial radiated field for different initial conditions, which we discuss in the next section below (Fig.~4). 

\subsection{Symmetric ensemble: time evolution of the spatial field}

We next discuss the time evolution of the spatial field in the far-field image plane. Figure 4 shows the absolute value of the observed radiated field, $\vert  \vc{E}_{obs} \vert$ in the image plane at different observation times, $t=0$, $10 \tau$, $20 \tau$, $30 \tau $, and $40 \tau$, respectively.  Figure~4 shows false-color plots in the $x-y$ plane, near the center of the ensemble: $-10 \lambda < x< 10 \lambda$ and $-10 \lambda < y< 10 \lambda$. Similar to Fig.~3, in Fig.~4, we show results for five different simulations which are chosen to be representative examples. At $t=0$, the observed field resembles a speckle pattern and there is very little spatial coherence. This is because of the initial random locations of the atoms in the ensemble. However, as the system evolves, because spatial correlations between the atoms are established (due to the dipole-dipole interaction), the emitted field forms a clear spatial pattern. 

The specific spatial pattern that is formed at later times is due to the system evolving into a dominant (or several prominent) subradiance modes. Depending on the initial randomly chosen location of the atoms, the specific spatial pattern that is formed in each simulation is different. However, while the specific pattern the system evolves in each simulation is different, long-range correlations, and therefore spatial coherence is established in each simulation.

\begin{figure}[h]
\vspace{-2cm}
\begin{center}
\includegraphics[width=18cm]{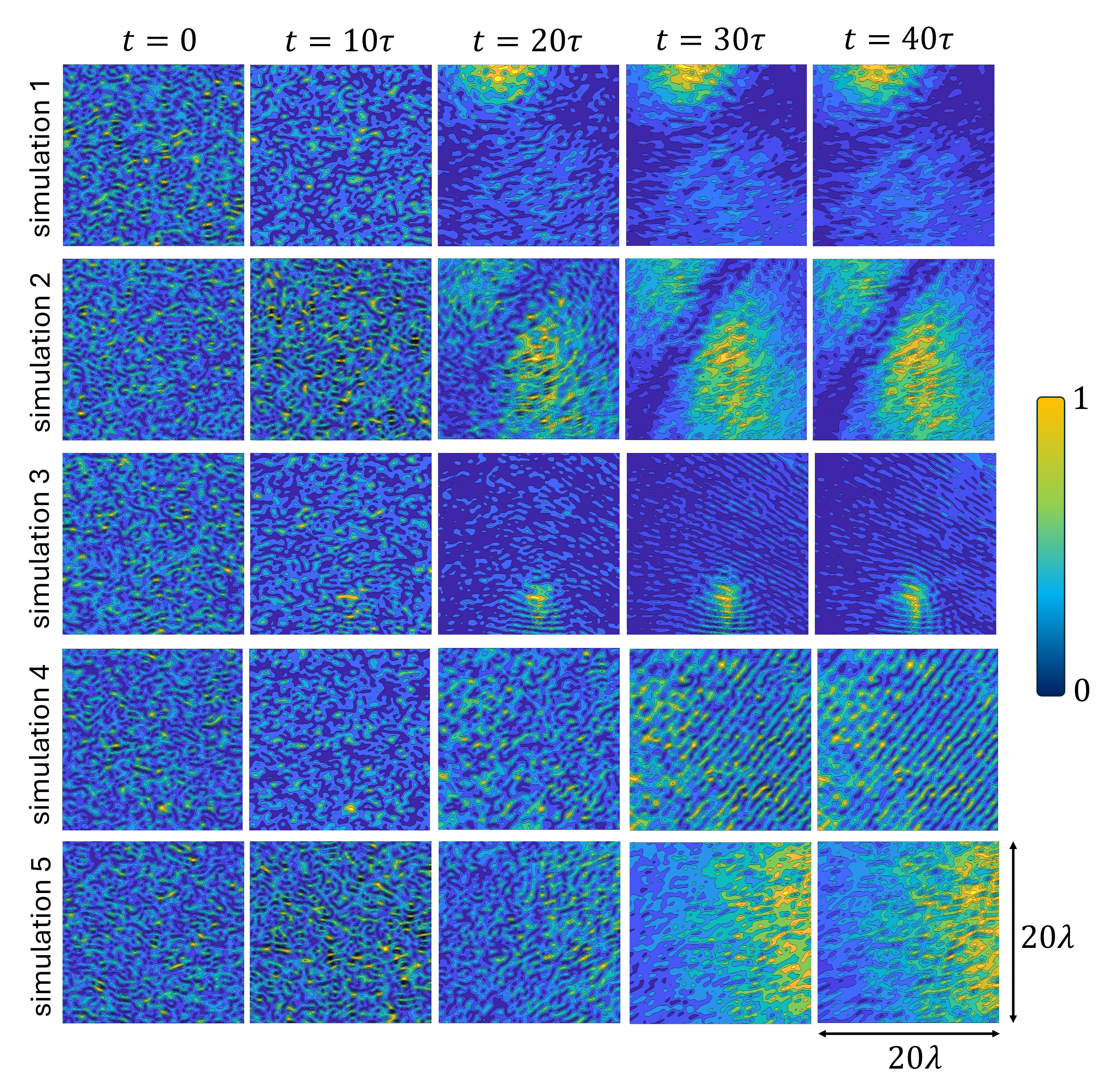}
\vspace{-1cm} 
\begin{singlespace}
\caption{\label{scheme} \small   False color plots of the absolute value of the radiated field, $\vert \vc{E}_{obs} \vert $ in the image plane at different observation times, $t=0$, $10 \tau$, $20 \tau$, $30 \tau $, and $40 \tau$, respectively. We show results for five different simulations which are chosen to be representative examples. The results are displayed in the $x-y$ plane, near the center of the ensemble: $-10 \lambda < x< 10 \lambda$ and $-10 \lambda < y< 10 \lambda$. At $t=0$, the observed field resembles a speckle pattern and there is very little spatial coherence. This is a result of the initial random locations of the atoms in the ensemble. However, as the system evolves, because spatial correlations between the atoms are established (due to the dipole-dipole interaction), the emitted field forms a clear spatial pattern. }
\end{singlespace}
\end{center}
\vspace{-0.2cm}
\end{figure}

\begin{figure}[h]
\vspace{-0.5cm}
\begin{center}
\includegraphics[width=9cm]{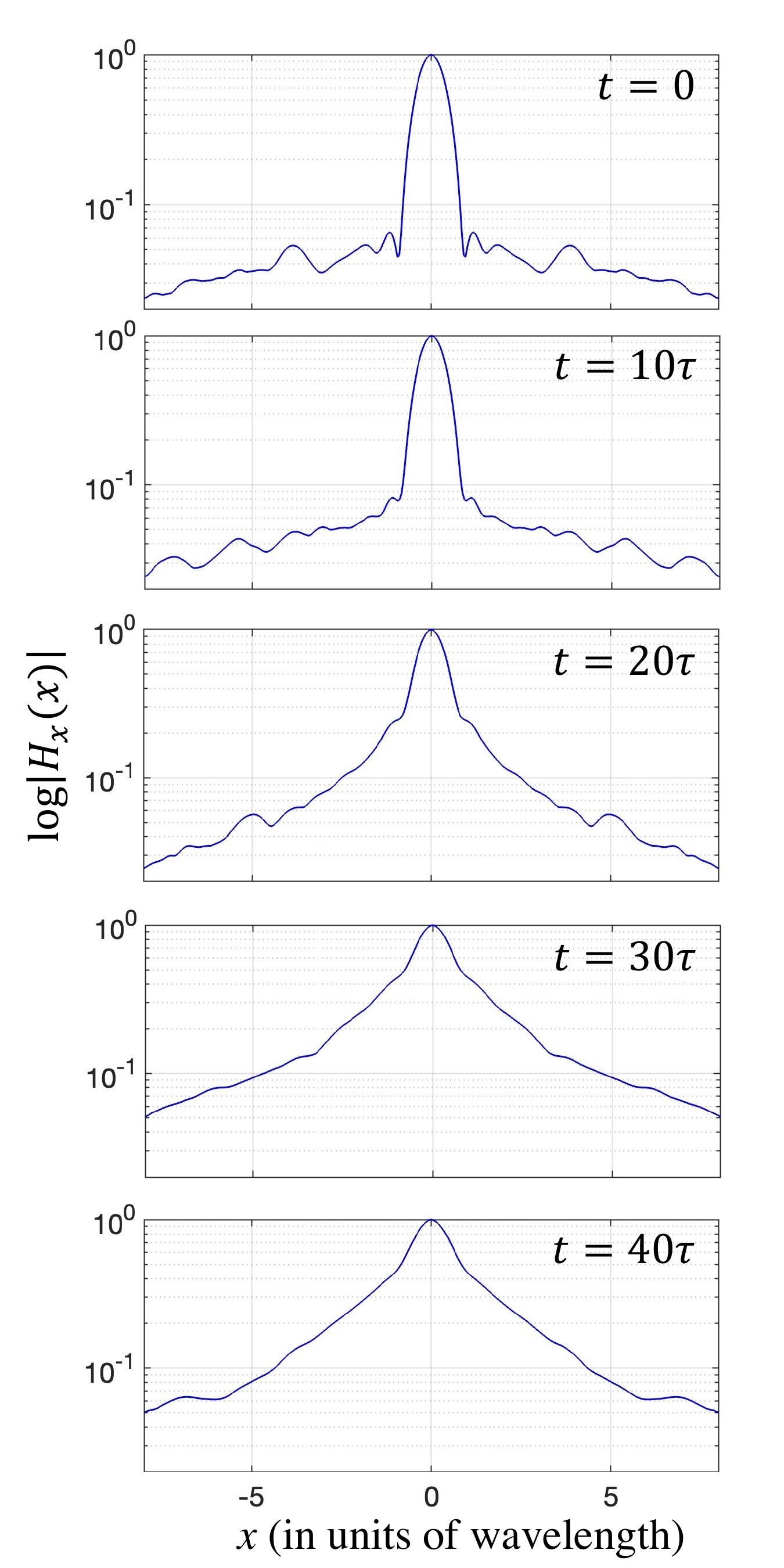}
\vspace{-0.2cm} 
\begin{singlespace}
\caption{\label{scheme} \small  $\log{ \vert H_x(x) \vert}$, at observation times of $t=0$, $10 \tau$, $20 \tau$, $30 \tau $, and $40 \tau$, respectively. As expected from the images of Fig.~4, initially at $t=0$ the cross-correlation function is sharply peaked, displaying very little spatial coherence. However, as the system evolves into the dominant subradiant modes, $H_x(x)$ acquires substantial width, displaying a large degree of spatial coherence.  
}
\end{singlespace}
\end{center}
\vspace{-0.cm}
\end{figure}

\subsection{Symmetric ensemble: time evolution of the cross-correlation function}

We proceed with a more quantitative analysis of the established spatial coherence. Because of the variation of the results due to initial random distribution of the atoms in the ensemble, in what follows, we run our simulations 20 times, and display the average of the results over these 20 instances. Figure~5 shows the cross-correlation function, $H_x(x)$, as a function of the evolution time for the conditions identical to those that were used in Figs.~3 and 4 (i.~e. with $N=1500$ atoms and a cloud size of  $L_x= L_y = L_z = 40 \lambda$ in the three axis). Here, we plot the logarithm of the absolute value, $\log{ \vert H_x(x) \vert}$, at observation times of $t=0$, $10 \tau$, $20 \tau$, $30 \tau $, and $40 \tau$, respectively.  As expected from the images of Fig.~4, initially (at $t=0$) the cross-correlation function is sharply peaked, displaying very little spatial coherence. However, as the system evolves into the dominant subradiant modes, $H_x(x)$ acquires substantial width, displaying a large degree of spatial coherence.

\begin{figure}[h]
\vspace{0.5cm}
\begin{center}
\includegraphics[width=18cm]{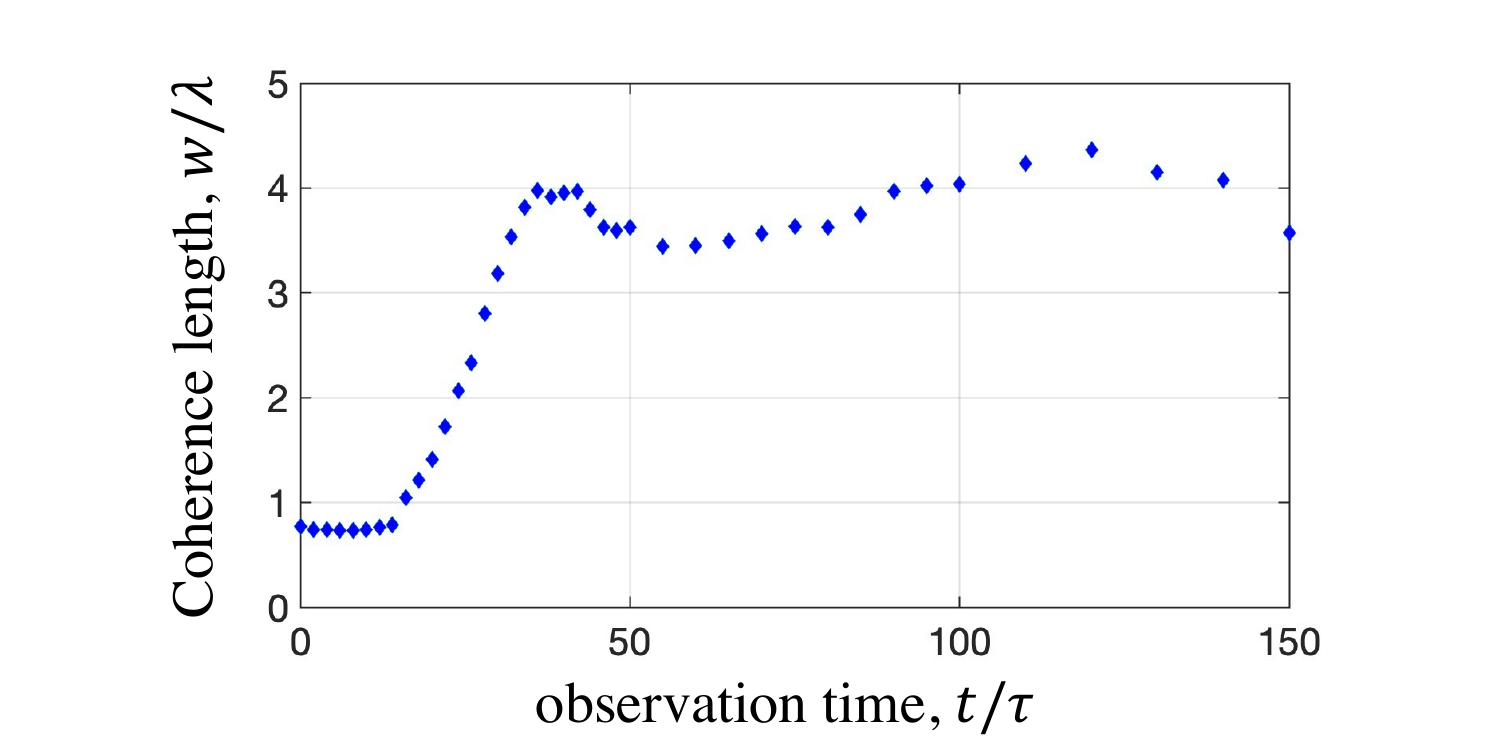}
\vspace{-0.2cm} 
\begin{singlespace}
\caption{\label{scheme} \small   The width of the cross-correlation function, $w$ (in units of wavelength) as a function of the evolution time. Up until about $t=10 \tau$, there is not much change in the coherence length $w$ since many subradiant modes continue to contribute to the emission. After about $t \sim 10 \tau$, there is a sharp rise in the coherence length, which reaches its peak value at about $t=40 \tau$ and then stabilizes.}
\end{singlespace}
\end{center}
\vspace{-0.cm}
\end{figure}

In Fig.~6, we plot the width of the cross-correlation function, $w$ (in units of wavelength), as a function of the evolution time. Up until about $t=10 \tau$, there is not much change in the coherence length since many collective eigenstates continue to contribute to the emission. After about $t \sim 10 \tau$, there is a sharp rise in the coherence length, which reaches its peak value at about $t=40 \tau$ and then stabilizes. This is because at about  $t=40 \tau$ the system has evolved into a few prominent subradiant modes, which are the main contributors to the emission pattern afterwards.

We note that in the simulations that we have discussed above, the observed time-scale for establishing the spatial coherence is rather long ($t > 10 \tau$). In contrast, in our recent experiment, the data was collected in a shorter time window of $0 < t < 5 \tau$ after the excitation laser was switched off \cite{davidexp}. This discrepancy is likely due to the single-excited subspace assumption of the numerical results that we have presented here. The experiment was performed in the strong excitation regime, and in this regime we would expect the time-scales for establishing the spatial coherence to get shorter. This is because, for higher excitation subspaces, the spread of the eigenvalues of the Hamiltonian (and therefore the spread of the decay rates of the subradiant modes) is larger.  An extension of the above numerical results to the strong excitation regime where the wavefunction explores a large portion of the Hilbert space is one future direction. One approach along this direction would be to first study doubly and triply excited subspaces, with dimensions of $N^2/2$ and $N^3/6$, respectively. It may then be possible to extrapolate the results to the full Hilbert space.

\subsection{Symmetric ensemble: the figure-of-merit (FoM) for spatial coherence in collective spontaneous emission}

We next proceed with a detailed discussion of the relevant figure-of-merit (FoM) for the ensemble that determines the degree of spatial coherence. We first note again that neither the density, nor the optical depth of the ensemble is a good FoM for the observed physics. As we mentioned above, for the numerical simulations of Figs.~3-6,  the ensemble is rather dilute, with an atomic density of $n=N/(L_x \times L_y \times L_z)=0.023/\lambda^3$, and is also optically thin, with an O.~D.$= n \sigma L_x =n \frac{\lambda^2}{2 \pi} L_x = 0.15$.  For the spatial coherence established in collective spontaneous emission, it is critical that the system is dilute with a low density. This is because, subradiant states that have long-range correlations across the ensemble are responsible for the observed coherence. At high densities, near-field interactions start to dominate and the eigenvectors of the Hamiltonian (the subradiant modes) mostly involve local excitations (i.e., modes with spatial correlations that are locally confined and that do not extend to the whole ensemble). 

\begin{figure}[h]
\vspace{-0.5cm}
\begin{center}
\includegraphics[width=18cm]{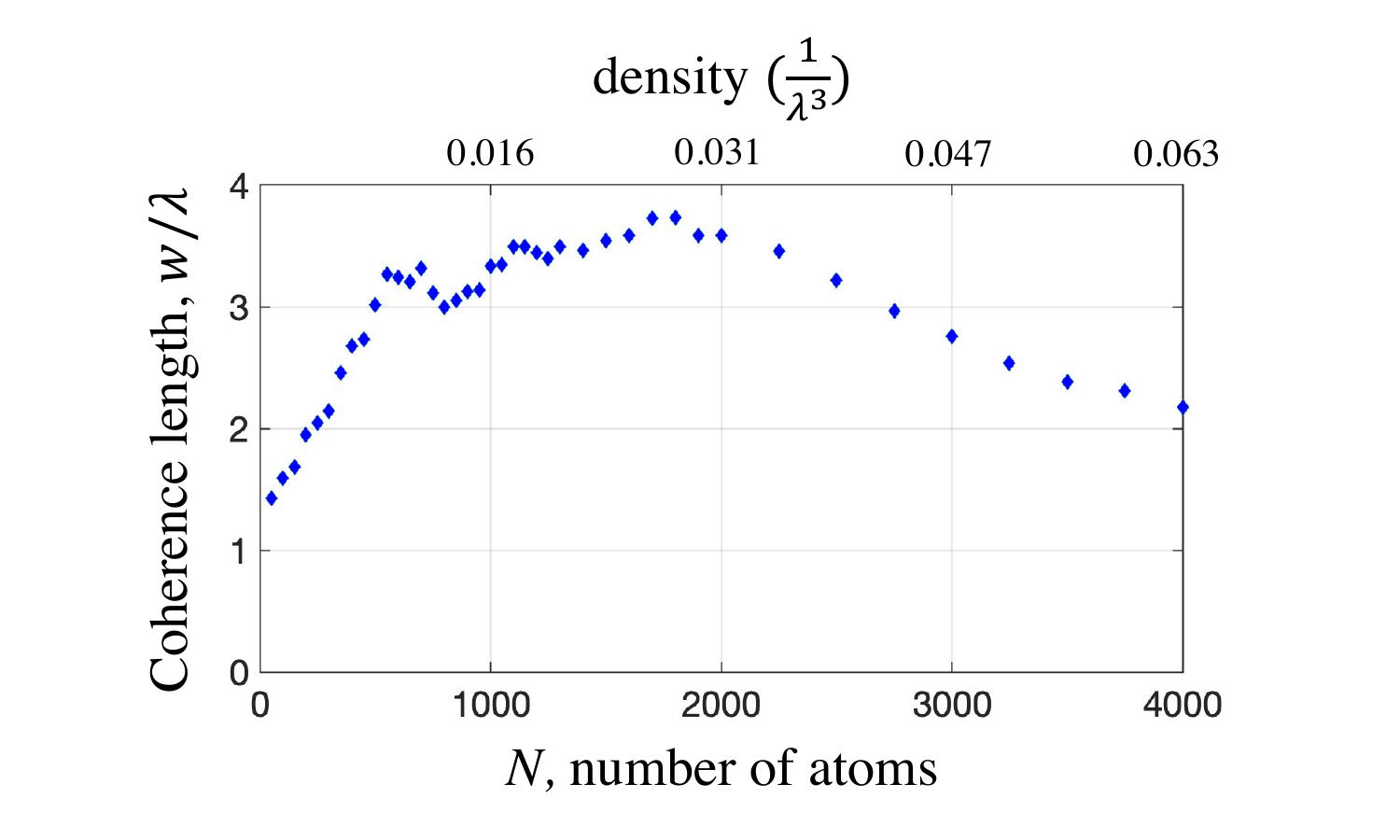}
\vspace{-1cm} 
\begin{singlespace}
\caption{\label{scheme} \small  The observed coherence length as a function of the $N$, the number of atoms in the ensemble. Here, the cloud size is the same as the simulations of Figs.~3-6 of above with $L_x= L_y = L_z = 40 \lambda$ and the observation time is fixed to $t=40 \tau$. As the atom number increases, there is an initial increase in the spatial coherence length. However, the established coherence reaches a plateau near $N=1500$ and drops with further increase in the number of atoms. This drop is due to the near-field interactions starting to dominate and the eigenvectors making a transition from being global excitations across the ensemble to local excitations. 
}
\end{singlespace}
\end{center}
\vspace{-0.5cm}
\end{figure}

Figure~7 shows this result. Here, the cloud size is the same as the simulations that we have discussed above with $L_x= L_y = L_z = 40 \lambda$. Furthermore, we fix the observation time to $t=40 \tau$, and plot the observed coherence length as a function of $N$, the number of atoms in the ensemble. As the atom number increases, there is an initial increase in the spatial coherence length. However, the established coherence reaches a plateau near $N=1500$ and drops with further increase in the number of atoms. This drop is due to the near-field interactions starting to dominate and the eigenvectors making a transition from being global excitations across the ensemble to local excitations.

\begin{figure}[tbh]
\vspace{0cm}
\begin{center}
\includegraphics[width=17cm]{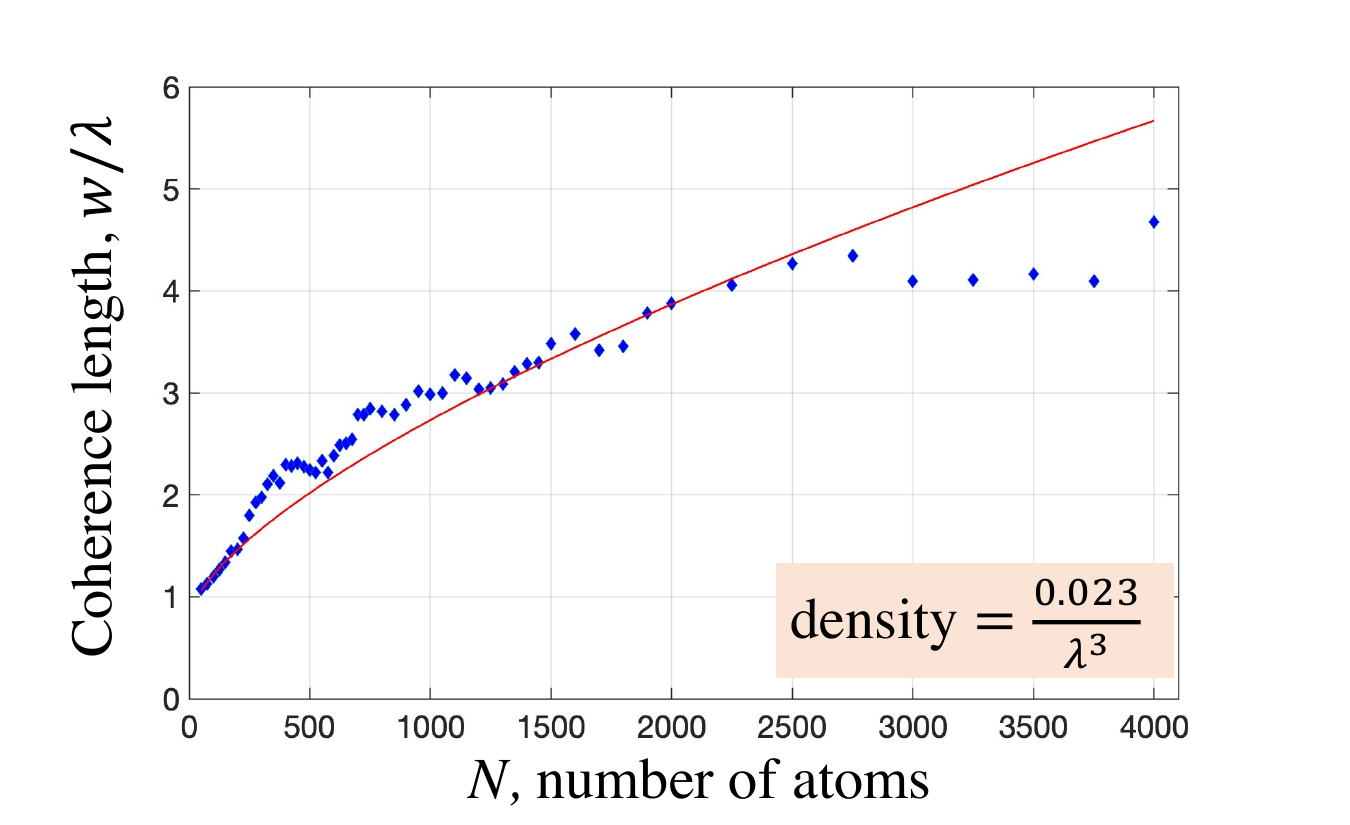}
\vspace{-0.5cm} 
\begin{singlespace}
\caption{\label{scheme} \small   The coherence length $w$ as a function of the number of atoms, $N$, at a fixed observation time of $t=40 \tau$.  We keep the density of the ensemble fixed at $n = 0.023 /\lambda^3$  and change the volume of the symmetric sample from a size of $12.8\lambda \times 12.8 \lambda \times 12.8\lambda$ to $55.5\lambda \times 55.5 \lambda \times 55.5\lambda$. As the size of the sample is changed, the number of atoms in the ensemble is varied from $N=50$ to $N=4000$ (so that the density is kept constant). The solid line is a fit to the numerical data using Eq.~(14), with a single fitting parameter of $\xi=0.45$. The model fits the numerical calculation reasonably well, with significant deviations especially at high atom numbers.  }
\end{singlespace}
\end{center}
\vspace{-0.3cm}
\end{figure}

In our recent experiment, we have observed and qualitatively explained our experimental results using the FoM$\sim \frac{N}{(L/\lambda)}$. This result is consistent with the width of the eigenvalue distribution of the exchange Hamiltonian in the large number of atoms limit, $N \rightarrow \infty$, which is given by \cite{ben}:
\begin{eqnarray}
\sigma_{\hat{H}}= \frac{1}{4 \pi} \sqrt{\left( \pi + \frac{29}{12} \right)} \frac{N}{(L/\lambda)} \quad .
\end{eqnarray}

Motivated by these results, we hypothesize that the spatial coherence length that is achieved during collective decay after the system evolves into dominant subradiant modes is given by:
\begin{eqnarray}
w=w_{u} \left(1 + \xi \sigma_{\hat{H}} \right)  = w_{u}\left(1 + \xi \frac{1}{4 \pi} \sqrt{\left( \pi + \frac{29}{12} \right)} \frac{N}{(L/\lambda)} \right) \quad . 
\end{eqnarray}

Here, as we will discuss below, the quantity $\xi$ is a fitting parameter with a numerical value of order unity and $w_{u}$ is the spatial coherence length that is observed when there are no dipole-dipole correlations in the system. The main idea behind Eq.~(14) is that emission from a single uncorrelated atom  produces a speckle with a width of $w_{u}$. The width of the eigenvalue distribution roughly is a measure of the average number of correlated atoms in each of the subradiant eigenmodes. As a result, when correlations are established, we would expect the uncorrelated width to go up by a factor proportional to $\sigma_{\hat{H}}$.  

To test this hypothesis, we have run simulations where we keep the density of the ensemble fixed at $n = 0.023 /\lambda^3$ (which is the point where maximum coherence length is observed from Fig.~7), and change the volume of the symmetric sample from a size of $12.8\lambda \times 12.8 \lambda \times 12.8\lambda$ to $55.5\lambda \times 55.5 \lambda \times 55.5\lambda$. As the size of the sample is changed, the number of atoms in the ensemble is varied from $N=50$ to $N=4000$ (so that the density is kept constant at the mentioned value). In Fig.~8, we plot the coherence length $w$ as a function of the number of atoms, $N$, again at a fixed observation time of $t=40 \tau$. The solid line is a fit to the numerical data using Eq.~(14), with the single fitting parameter $\xi$. The best fit is obtained with $\xi=0.45$ (solid red line) which fits the data points reasonably well, with significant deviations especially at high atom numbers. 

\begin{figure}[tbh]
\vspace{0cm}
\begin{center}
\includegraphics[width=15cm]{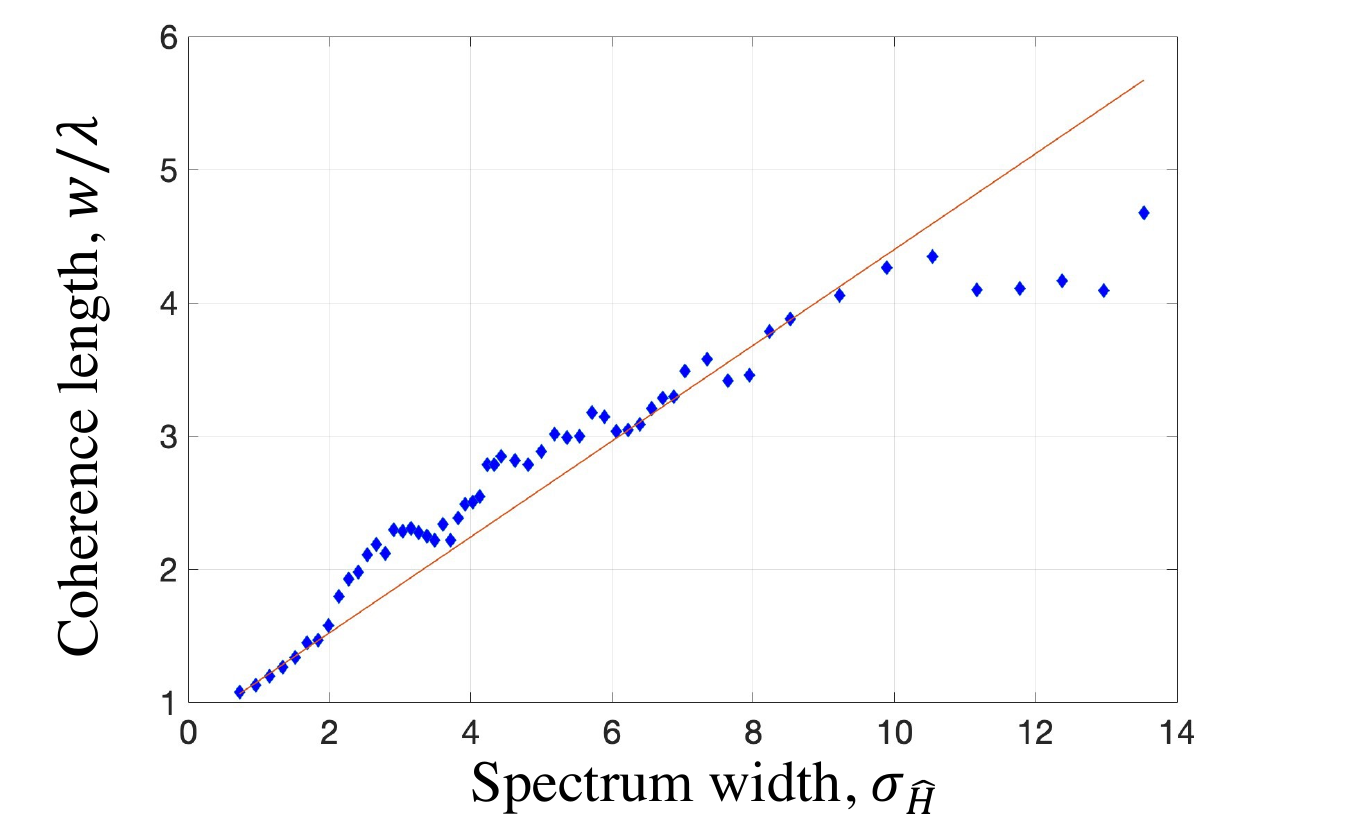}
\vspace{-0cm} 
\begin{singlespace}
\caption{\label{scheme} \small The numerically calculated coherence length as a function of the width of the eigenvalue spectrum $\sigma_{\hat{H}}= \frac{1}{4 \pi} \sqrt{\left( \pi + \frac{29}{12} \right)} \frac{N}{(L/\lambda)}$, displaying a near-linear dependence. The solid line is the same best fit using Eq.~(14) with again $\xi=0.45$.   }
\end{singlespace}
\end{center}
\vspace{-0.3cm}
\end{figure}

Figure~9 shows the same numerical results that are displayed in Fig.~8, but plotted as a function of the eigenvalue spectrum width, $\sigma_{\hat{H}}$, of Eq.~(13). The solid line is the same best fit of the form Eq.~(14) with again $\xi=0.45$. The fit is again reasonable with significant deviations of the data points from the fit at higher values of the spectrum width (corresponding to higher atom numbers). In the future, by extending our simulations to atom numbers in the $10^4$ to $10^5$ range, it may be possible to quantitatively analyze the nature of the discrepancy at higher atom numbers. 

\subsection{Numerical simulation results for asymmetric ensembles}

A detailed quantitative study of spatial coherence in collective spontaneous emission with different shapes of the atomic ensemble is beyond the scope of this paper. In this section, we present a preliminary investigation of spatial coherence in asymmetric clouds. For this purpose, we have numerically simulated two cases: in the first case, the atomic cloud is elongated along the $x$ axis, while in the second case the ensemble is elongated along the $z$ axis. For a clear comparison with the symmetric-cloud investigations of Figs.~4-7, we have kept the volume of the ensemble the same. Specifically, in Figure~10, we consider an ensemble with a size of $80 \lambda \times 40 \lambda \times 20 \lambda$, which is elongated along the $x$ axis. In Fig.~10(a), we plot the spatial coherence length as a function of observation time when the atom number is fixed at $N=1500$. In Fig.~10(b), we fix the observation time to $t=40 \tau$ and plot the coherence length as a function of the number of atoms in the ensemble. The results are qualitatively similar to the symmetric case (Figs.~6 and 7), with the overall degree of spatial coherence being about $25 \%$ lower. 

\begin{figure}[tbh]
\vspace{-0.5cm}
\begin{center}
\includegraphics[width=15cm]{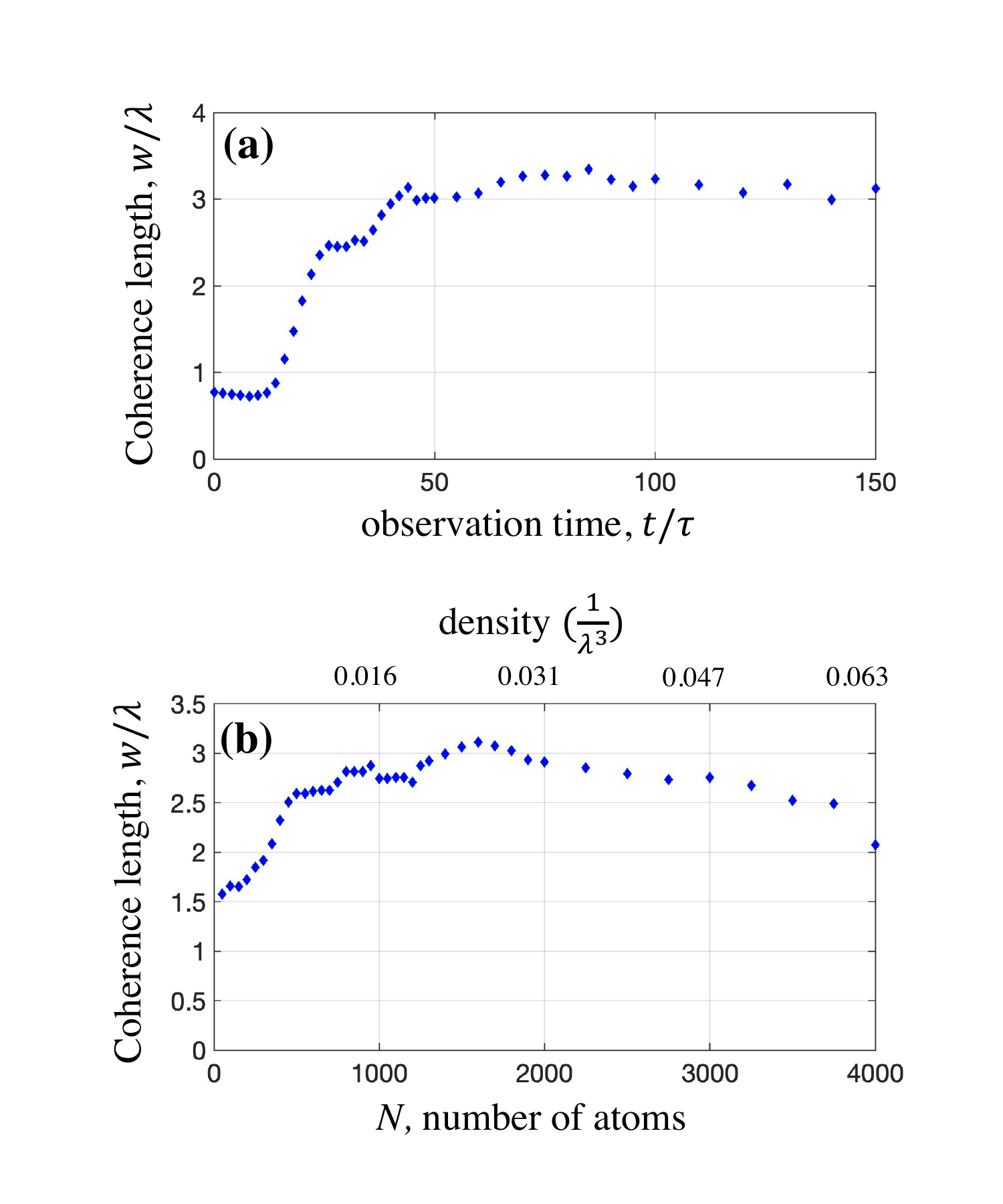}
\vspace{-1cm} 
\begin{singlespace}
\caption{\label{scheme} \small   Numerical results for an ensemble with a size of $80 \lambda \times 40 \lambda \times 20 \lambda$ (elongated along the $x$ axis). The volume of the ensemble is the same as the symmetric cloud investigations of Figs.~4-7. (a) The spatial coherence length as a function of observation time with the atom number fixed at $N=1500$. (b) We fix the observation time to $t=40 \tau$ and plot the coherence length as a function of the number of atoms in the ensemble. The results are qualitatively similar to the symmetric case (Figs.~6 and 7), with the overall degree of spatial coherence being about $25 \%$ lower. }
\end{singlespace}
\end{center}
\vspace{-0.5cm}
\end{figure}

In Figure~11, we consider an ensemble with a size of $20 \lambda \times 40 \lambda \times 80 \lambda$, which again has the same volume as the symmetric-ensemble simulations of above, but which is elongated along the $z$ axis. In Fig.~11(a), we plot the spatial coherence length as a function of observation time when the atom number is fixed at $N=1500$. In Fig.~11(b), we fix the observation time to $t=40 \tau$ and plot the coherence length as a function of the number of atoms in the ensemble. Again the results are qualitatively similar to the symmetric case (Figs.~6 and 7). However, in this case, the overall degree of spatial coherence is about $50\%$ higher. A detailed study of the spatial coherence in collective spontaneous emission with different shapes of the atomic ensemble is left for future work. 

\begin{figure}[tbh]
\vspace{-0.3cm}
\begin{center}
\includegraphics[width=15cm]{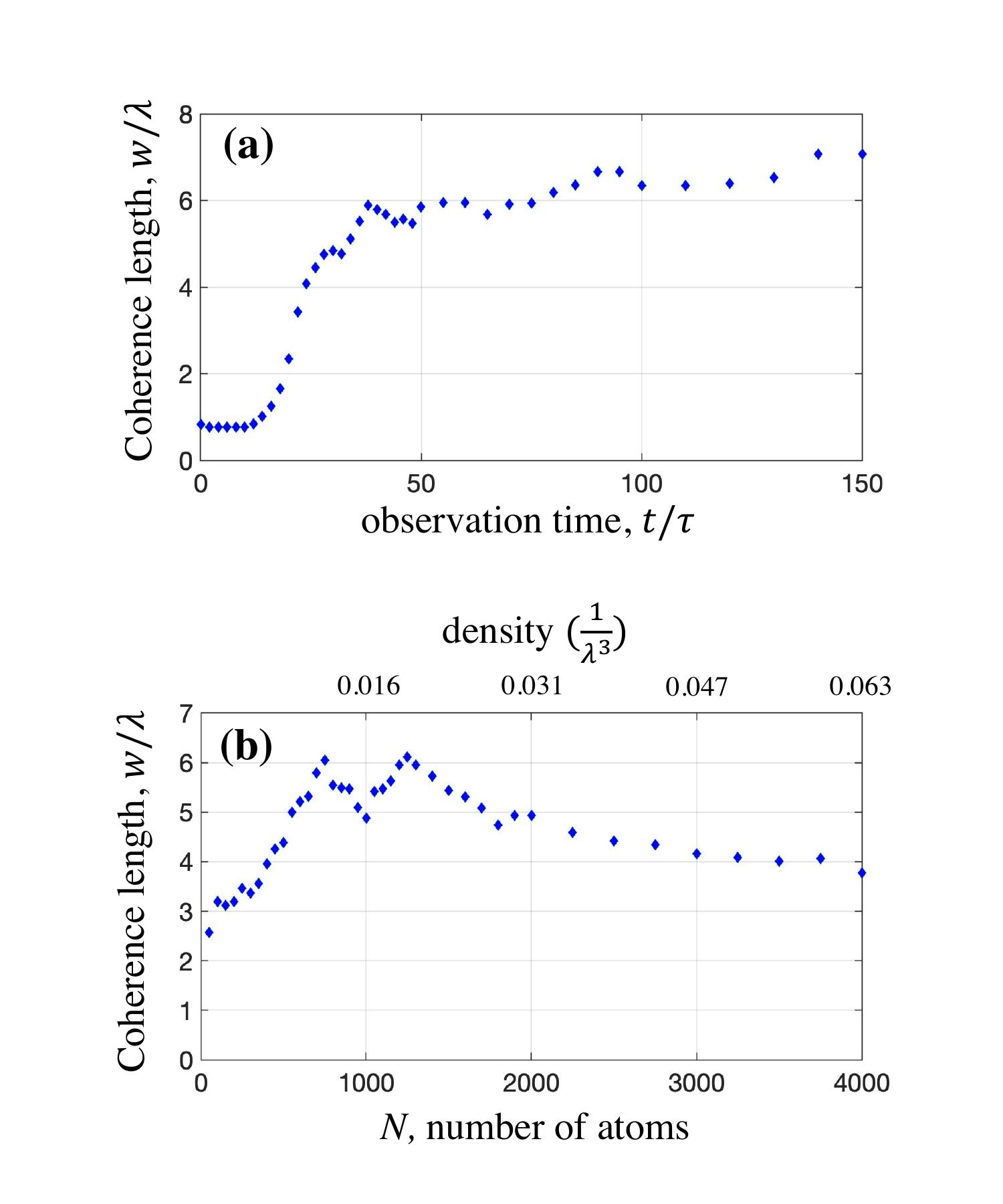}
\vspace{-1cm} 
\begin{singlespace}
\caption{\label{scheme} \small   Numerical results for an ensemble with a size of $20 \lambda \times 40 \lambda \times 80 \lambda$ (elongated along the $z$ axis). (a) The spatial coherence length as a function of observation time when the atom number is fixed at $N=1500$. (b) We fix the observation time to $t=40 \tau$ and plot the coherence length as a function of the number of atoms in the ensemble. Again, the results are qualitatively similar to the symmetric case (Figs.~6 and 7), with the overall degree of spatial coherence about $50 \%$ higher. }
\end{singlespace}
\end{center}
\vspace{-0.cm}
\end{figure}

\section{Conclusions}

In conclusion, we have presented a numerical study of the spatial coherence of light in collective spontaneous emission. As we mentioned above, the spatial coherence of light that is produced in collective decay in dilute atomic clouds is distinctly different from the coherence of a laser. It does not rely on population inversion which is critically required for the operation of a laser. Furthermore, in lasers, the phasing of the atomic dipoles through stimulated emission is necessary. In contrast, in dilute clouds experiencing collective decay, subradiance (out-of-phase superpositions) is the dominant mechanism that establishes correlations between the atomic dipoles, and therefore the spatial coherence. Finally, the spatial coherence in collective spontaneous emission is governed by the FoM of  $\sim\frac{N}{(L/\lambda)}$ instead of the gain-length product (or the optical depth) of the ensemble. 

In our simulations, we have assumed weak excitation of the ensemble and restricted the Hilbert space to the single-excited subspace. An extension of these results to the strong excitation regime where the wavefunction explores a large portion of the Hilbert space is one future direction. One approach along this direction would be to first study doubly and triply excited subspaces, with dimensions of $N^2/2$ and $N^3/6$, respectively. It may then be possible to extrapolate the results to the full Hilbert space. As we mentioned above, we expect the time-scales for establishing the spatial coherence to get shorter, as the system moves into subspaces with a higher dimension. This is because, for higher excitation subspaces, the spread of the eigenvalues of the Hamiltonian (and therefore the spread of the decay rates of the subradiant modes) is larger \cite{ben}.

On the experimental side, the spatial patterns that are established during collective spontaneous emission (as shown in Fig.~3) are directly related to the structure of the subradiant eigenvectors of the Hamiltonian. It would be very exciting to experimentally observe these spatial patterns. As we discussed above, when the ensemble is disordered, the radiated field evolves into different spatial patterns due to random initial conditions of the atoms. It may be possible to steer the evolution by imposing some initial structure to the radiating cloud, for example, using an ordered array of atoms. We note that in the strong excitation regime, even finding a single-eigenvector of the exchange Hamiltonian is likely an NP-hard problem due to the exponentially large dimension of the Hilbert space \cite{ising1,ising2}. By imaging the observed spatial patterns in the strong excitation regime (using, for example, a photon counter array), the experiments could give information regarding the structure of the subradiant eigenmodes in the exponentially-large Hilbert space. These subradiant eigenstates cannot be efficiently computed using classical means. 

In our simulations, we have assumed initial excitation of the atomic ensemble by a laser beam, which was the case in our recent experiment \cite{davidexp}. It is likely that these results will extend to the case when the excitation is incoherent, since the dipole-dipole correlations that are established (and that are responsible for the established spatial coherence) do not rely on the phase-coherence of the initial laser pulse. Simulating collective decay under the conditions of incoherent excitation to the excited level is another future direction. 

Finally, we pose the following question: what are the quantum statistics of the emitted photons during collective spontaneous emission in the regime that we have discussed? Both theoretical and experimental studies of the quantum mechanical statistics of the emitted photons is yet another future direction. Experimentally, the fluctuations in the two quadratures of the spontaneously emitted light can be measured by beating the radiation coming from the ensemble with a local oscillator in a homodyne detection set-up \cite{mandel}. 

\section{Acknowledgements}
We thank Francis Robicheaux from Purdue University for many helpful discussions regarding our prior experimental work. This work was supported by the National Science Foundation (NSF) Grant No. 2016136 for the QLCI center Hybrid Quantum Architectures and Networks (HQAN), NSF Grant No. 2308818 from the AMO-Experiment program, and also by the University of Wisconsin-Madison, through the Vilas Associates award.

\newpage
\section{Appendix A: Derivation of the far field radiation pattern}
In this Appendix, we provide a derivation of the far-field electric field expressions from the ensemble, Eqs.~(10) and (11) of above. We start with the classical expression for the radiated electric field from a source with the current density $\vc{J}(\vc{x},t)$: 
\begin{equation}
    \vc{E}_{\text{rad}}(\vc{x},t)=\frac{1}{4\pi \epsilon_0 c^2}\int \frac{(\Dot{\vc{J}}(\vc{x}',t-r/c)\times\vc{r})\times\vc{r}}{r^3}d^3x', \quad \vc{r}=\vc{x}-\vc{x}' \quad . 
\end{equation}
For the quantum treatment of the radiation from the ensemble, we start by defining the source operator field (while working in the Heisenberg picture): 
\begin{equation}
    \qp{\vc{J}}(\vc{x}',t)=\sum_{i=1}^{N}\delta^{(3)}_{\varepsilon}(\vc{x}'-\vc{x}_{i})\frac{d}{dt}\hat{\vc{P}}_i(t) \quad . 
\end{equation}
\noindent Here, $\delta^{(3)}_{\varepsilon}$ is a three dimensional coarse-grained delta function such that the support length $\varepsilon$ is much greater than the atomic scale $r_0$ (Bohr radius) but is still considerably smaller than the wavelength $\lambda$ of the emitted light \cite{haroche}. The definition of the dipole moment operator for atom $i$, $\qp{\vc{P}}_{i}(t)$ (again in the Heisenberg picture) is: 
\begin{equation}
    \qp{\vc{P}}_i(t)=\bv{\epsilon}_a \qp{\mu}_i(t), \quad \qp{\mu}_i(t)=\overbrace{\qp{1}\otimes\dots\otimes}^{i-1\text{ terms}}e^{i\qp{H}_{eff} t}\;\qp{\mu}\;e^{-i\qp{H}_{eff}t}\otimes\dots\otimes\qp{1}  
\end{equation}
Here, the operators $\qp{1}$ are identity operators for all other atoms (other than atom $i$.) Using Eqs.~(16) and (17) of above, the quantum mechanical analog of the classical radiated field is:
\begin{align}
    \qp{\vc{E}}_{\text{rad}}(\vc{x},t)&=\frac{1}{4\pi\epsilon_0c^2}\int \sum_{i=1}^{N}\delta^{(3)}_{\varepsilon}(\vc{x}'-\vc{x}_{i})\left[\frac{d^2}{dt^2}\hat{\mu}_i(t)\right]_{t-r/c}\frac{(\bv{\epsilon}_a\times\vc{r})\times\vc{r}}{r^3}d^3x' \nonumber \\
    &\approx\frac{1}{4\pi\epsilon_0c^2}\sum_{i=1}^{N}\left[\frac{d^2}{dt^2}\hat{\mu}_i(t)\right]_{t-r_i/c}\frac{(\bv{\epsilon}_a\times\vc{r}_i)\times\vc{r}_i}{r_i^3}, \quad \vc{r}_i=\vc{x}-\vc{x}_i\quad . 
\end{align}
Using Eq.~(18), we can calculate the average radiation field at the point of observation due to the collective spontaneous emission process: 
\begin{equation}
    \vc{E}_{\text{rad}}(\vc{x},t)=\bra{\Psi(0)}\qp{\vc{E}}_{\text{rad}}(\vc{x},t)\ket{\Psi(0)}  \quad . 
\end{equation}
Here, $\ket{\Psi(0)}$ is the initial wavefunction of the system after the excitation laser is switched-off: 
\begin{equation}
    \ket{\Psi(0)}=\cos(\theta/2)\ket{00\dots0}+\sin(\theta/2)\ket{\psi(0)} \quad . 
\end{equation}
The state  $\ket{00\dots0}$ is the state where all the atoms are in their ground state, which is the collective atomic state before the excitation laser is applied. The excitation laser rotates the system to a superposition  of the collective ground state and the single-excited subspace of above, with a rotation angle $\theta$ which is given by: 
\begin{equation}
    \tan{\theta}=\frac{|\Omega_\ell|}{\Delta},\quad \Omega_{\ell}=\frac{E_{\text{laser}}}{2}\bra{1}\hat{\mu}\ket{0} \quad . 
\end{equation}
Here, the quantity $\Delta$ is the detuning of the excitation laser beam from the atomic transition frequency.  We then have:
\begin{align}
    \vc{E}_{\text{rad}}(\vc{x},t)&=\frac{1}{4\pi\epsilon_0c^2}\sum_{i=1}^{N}\left[\frac{d^2}{dt^2}\bra{\Psi(0)}\hat{\mu}_i(t)\ket{\Psi(0)}\right]_{t-r_i/c}\frac{(\bv{\epsilon}_a\times\vc{r}_i)\times\vc{r}_i}{r_i^3}\\
    &=\frac{1}{4\pi\epsilon_0c^2}\sum_{i=1}^{N}\left[\frac{d^2}{dt^2}\bra{\Psi(t)}\hat{\mu}_i\ket{\Psi(t)}\right]_{t-r_i/c}\frac{(\bv{\epsilon}_a\times\vc{r}_i)\times\vc{r}_i}{r_i^3} \quad . 
\end{align}
Using Eq.~(7) of above, we then have: 
\begin{align}
&\bra{\Psi(t)}\hat{\mu}_i\ket{\Psi(t)}=\sin\theta\Re\bra{00\dots0}\qp{\mu}_i\ket{\psi(t)}=\sin\theta\Re\sum_{q}e^{-i\omega_a t}e^{-\Gamma t/2}c_q(t)\bra{00\dots0}\qp{\mu}_i\ket{q}\\
&=\sin\theta\Re e^{-i\omega_a t}e^{-\Gamma t/2}c_i(t)\bra{0}\qp{\mu}\ket{1}=\frac{|\Omega_{\ell}|}{\sqrt{|\Omega_\ell|^2+\Delta^2}}\Re e^{-i\omega_a t}e^{-\Gamma t/2}c_i(t)\bra{0}\qp{\mu}\ket{1}
\quad . 
\end{align}
Substituting the relation given in Eq.~(25) back into Eq.~(23,), differentiating only the highly oscillatory part $e^{-i\omega_a t}$ (with $\;\omega_a/2\pi\sim 10^{15}\gg \Gamma$) twice, ignoring all other subleading terms and retarded time for long time scales of $\tau$ because $\Gamma z_{obs}/c\ll1$, we obtain: 
\begin{align}
    \vc{E}_{\text{rad}}(\vc{x},t)&=\frac{|\Omega_{\ell}|}{\sqrt{|\Omega_\ell|^2+\Delta^2}}\Re\frac{\omega_a^2}{4\pi\epsilon_0c^2}\sum_{i=1}^{N}\frac{e^{ik_ar_i}}{r_i}[(\hat{r}_i\times\bra{0}\qp{\mu}\ket{1}\bv{\epsilon}_a)\times\hat{r}_i]e^{-i\omega_a t}e^{-\Gamma (t-r_i/c)/2}c_i(t-r_i/c)+ O(\omega_a)\nonumber \\
    &\approx \frac{|\Omega_{\ell}|}{\sqrt{|\Omega_\ell|^2+\Delta^2}}\Re\sum_{i=1}^{N}\overbrace{\frac{k_a^3}{4\pi\epsilon_0}\frac{\exp{ik_ar_i}}{k_a r_i}[(\hat{r}_i\times\vc{p})\times\hat{r}_i]e^{-i\omega_a t}e^{-\Gamma t/2}c_i(t)}^{\vc{E}_i(\vc{x},t)} \quad . 
\end{align}


\newpage
\begin{references}

\bibitem{dicke} R. H. Dicke, Coherence in Spontaneous Radiation Processes, Phys. Rev. {\bf 93}, 99 (1954).

\bibitem{haroche} M. Gross and S. Haroche, Superradiance: An Essay on the Theory of Collective Spontaneous Emission, Phys. Rep. {\bf 93}, 301 (1982).

\bibitem{eberly} L. Allen and G. H. Eberly, Optical Resonance and Two-Level Atoms, Dover Publications (1987). 

\bibitem{scully1} M. O. Scully and A. A. Svdzinsky, The Super of Superradiance, Science {\bf 325}, 1510 (2009).

\bibitem{yelin} H. Ma, O. Rubies-Bigorda, and S. F. Yelin, Superradiance and Subradiance in a Gas of Two-level Atoms, arXiv:2205.15255 [quant-ph] (2022). 

\bibitem{francis} F. Robicheaux and D. A. Suresh, Beyond lowest order mean-field theory for light interacting with atom arrays, Phys. Rev. A {\bf 104}, 023702 (2021). 

\bibitem{adams} R. J. Bettles, S. A. Gardiner, and C. S. Adams, Cooperative Eigenmodes and Scattering in One-Dimensional Atomic Arrays, Phys. Rev. A {\bf 94}, 043844 (2016). 

\bibitem{jenkins} S. D. Jenkins and J. Ruostekoski, Controlled Manipulation of Light by Cooperative Response of Atoms in an Optical Lattice, Phys. Rev. A {\bf 86}, 031602 (2012).

\bibitem {ritsch} H. Zoubi and H. Ritsch, Metastability and Directional Emission Characteristics of Excitons in 1D Optical Lattices, Europhys. Lett. {\bf 90}, 23001 (2010).

\bibitem{bachelard} C. E. Máximo,  R. Bachelard, F. E. A. dos Santos, and C. J. Villas-Boas, Cooperative Spontaneous Emission via Renormalization Approach: Classical Versus Semi-Classical Effects,  arXiv:1906.05719 (2020). 

\bibitem{petrov} D. F. Kornovan, A. S. Sheremet, and M. I. Petrov, Collective Polaritonic Modes in an Array of Two-Level Quantum Emitters Coupled to an Optical Nanofiber, Phys. Rev. B {\bf 94}, 245416 (2016).

\bibitem{zanthier} D. Bhatti, R. Schneider, S. Oppel and J. von Zanthier , Directional Dicke Subradiance with Nonclassical and Classical Light Sources, Phys. Rev. Lett. {\bf 120}, 1136 (2018).

\bibitem{agarwal} R. Wiegner, J. von Zanthier, and G. S. Agarwal, Quantum-interference-initiated Superradiant and Subradiant Emission from Entangled Atoms, Phys. Rev. A {\bf 84}, 023805 (2011).

\bibitem{reitz} M. Reitz, C. Sommer, and C. Genes, Cooperative Quantum Phenomena in Light-Matter Platforms, PRX Quantum {\bf 3}, 010201 (2022).


\bibitem{bloch} J. Rui, D. Wei, A. Rubio-Abadal, S. Hollerith, J. Zeiher, Dan M. Stamper-Kurn, C. Gross, and I. Bloch, A Subradiant Optical Mirror Formed by a Single Structured Atomic Layer, Nature {\bf 583}, 369 (2020).

\bibitem{an} J. Kim, D. Yang, S. Oh, and K. An, Coherent Single-Atom Superradiance, Science 359, 662 (2018). 

\bibitem{gauthier} J. A. Greenberg and D. J. Gauthier, Steady-state, Cavityless, Multimode Superradiance in a Cold Vapor, Phys. Rev. A {\bf 86}, 013823 (2012). 

\bibitem{kuga} Y. Yoshikawa, Y. Torii, and T. Kuga, Superradiant Light Scattering from Thermal Atomic Vapors, Phys. Rev. Lett. {\bf 94}, 083602 (2005). 

\bibitem{ions} R. G. DeVoe and R. G. Brewer, Observation of Superradiant and Subradiant Spontaneous Emission of Two Trapped Ions, Phys. Rev. Lett. {\bf 76}, 2049 (1996).

\bibitem{molecules} B. McGuyer, M. McDonald, G. Iwata et al., Precise Study of Asymptotic Physics with Subradiant Ultracold Molecules, Nature Phys. {\bf 11}, 32 (2015).

\bibitem{diamond1} C. Bradac, M. T. Johnsson, M. V. Breugel, et al., Room-temperature Spontaneous Superradiance From Single Diamond Nanocrystals, Nat. Commun. {\bf 8}, 1205 (2017).

\bibitem{diamond2} A. Angerer, K. Streltsov, T. Astner, T. et al., Superradiant Emission From Color Centers in Diamond, Nat. Phys. {\bf 14}, 1168 (2018).

\bibitem{superconducting} Z. Wang et al., Controllable Switching Between Superradiant and Subradiant States in a 10-qubit Superconducting Circuit, Phys. Rev. Lett. {\bf 124}, 013601 (2020).

\bibitem{feld}  N. Skribanowitz, I. P. Herman, J. C. MacGillivray, and M. S. Feld, Observation of Dicke Superradiance in Optically Pumped HF Gas, Phys. Rev. Lett. {\bf 30}, 309 (1973).

\bibitem{manassah} R. Friedberg, S. R. Hartmann, and J. T. Manassah, Frequency Shifts in Emission and Absorption by Resonant Systems of Two-level Atoms, Phys. Rep. {\bf 7}, 101 (1973).

\bibitem{kaiser1} W. Guerin, M. O. Araujo, and R. Kaiser, Subradiance in a Large Cloud of Cold Atoms, Phys. Rev. Lett. 116, 083601 (2016).

\bibitem{kaiser2} P. Weiss, M. O Araújo, R. Kaiser and W. Guerin, Subradiance and Radiation Trapping in Cold Atoms, New J. Phys. {\bf 20}, 063024 (2018).

\bibitem{kaiser3} T. Bienaime, N. Piovella, and R. Kaiser, Controlled Dicke Subradiance from a Large Cloud of Two-Level Systems, Phys. Rev. Lett. {\bf 108}, 123602 (2012). 

\bibitem{kaiser4} A. Cipris, N. A. Moreira, T. S. E. Santo, P. Weiss, C. J. Villas-Boas, R. Kaiser, W. Guerin, and R. Bachelard, Subradiance with Saturated Atoms: Population Enhancement of the Long-Lived States, Phys. Rev. Lett. {\bf 126}, 103604 (2021). 

\bibitem{kaiser5} P. Weiss, A. Cipris, M. O. Araujo, R. Kaiser, and W. Guerin, Robustness of Dicke subradiance against thermal decoherence, Phys. Rev. A {\bf 100}, 033833 (2019). 

\bibitem{browaeys1} Giovanni Ferioli, Antoine Glicenstein, Loic Henriet, Igor Ferrier-Barbut , and Antoine Browaeys, Storage and Release of Subradiant Excitations in a Dense Atomic Cloud, Phys. Rev. X {\bf 11}, 021031 (2021). 

\bibitem{browaeys2} A. Glicenstein, G. Ferioli, A. Browaeys, and I Ferrier-Barbut, From superradiance to subradiance: exploring the many-body Dicke ladder, Opt. Lett. {\bf 47}, 1541 (2022). 

\bibitem{browaeys3} G. Ferioli, A. Glicenstein, F. Robicheaux, R. T. Sutherland, A. Browaeys, and I. Ferrier-Barbut, Laser-Driven Superradiant Ensembles of Two-Level Atoms near Dicke Regime, Phys. Rev. Lett. {\bf 127}, 243602 (2021). 

\bibitem{dipto} D. Das, B. Lemberger, and D. D. Yavuz, Subradiance and Superradiance-to-Subradiance Transition in Dilute Atomic Clouds, Phys. Rev. A {\bf 102}, 043708 (2020).

\bibitem{davidexp} D. C. Gold, P. Huft, C. Young, A. Safari, T. G. Walker, M. Saffman, and D. D. Yavuz, Spatial Coherence of Light in Collective Spontaneous Emission, PRX Quantum {\bf 3}, 010338 (2022). 

\bibitem{ben} B. Lemberger and D. D. Yavuz, Effect of Correlated Decay on Fault-tolerant Quantum Computation, Phys. Rev. A {\bf 96}, 062337 (2017).

\bibitem{siegman} A. E. Siegman, Lasers, University Science Books (1986).

\bibitem{yelin1} E. Shahmoon, D. S. Wild, M. D. Lukin, and S. F. Yelin, Cooperative Resonances in Light Scattering from Two-Dimensional Atomic Arrays, Phys. Rev. Lett. {\bf 118}, 113601 (2017).

\bibitem{yelin2} O. Rubies-Bigorda, S. Ostermann, and S. F. Yelin, Generating multi-excitation subradiant states in incoherently excited atomic arrays, arXiv:2209.00034 [quant-ph] (2022).

\bibitem{francis1} R. T. Sutherland and F. Robicheaux, Coherent Forward Broadening in Cold Atom Clouds, Phys. Rev. A {\bf 93}, 023407 (2016).  

\bibitem{francis2} D. A. Suresh and F. Robicheaux, Photon-induced atom recoil in collectively interacting planar arrays, Phys. Rev. A {\bf 103}, 043722 (2021).

\bibitem{francis3} F. Robicheaux, Theoretical study of early-time superradiance for atom clouds and arrays, Phys. Rev. A {\bf 104}, 063706 (2021). 

\bibitem{ritsch2} H. Zoubi and H. Ritsch, Lifetime and Emission Characteristics of Collective Electronic Excitations in Two-Dimensional Optical Lattices, Phys. Rev. A {\bf 83}, 063831 (2011).

\bibitem{ballantine} K. E. Ballantine and J. Ruostekoski, Quantum Single Photon Control, Storage, and Entanglement Generation with Planar Atomic Arrays, PRX Quantum {\bf 2}, 040362
(2021).

\bibitem{garcia} E. Sierra, S. J. Masson, and A. Ansejo-Garcia, Dicke Superradiance in Ordered Lattices: Dimensionality Matters, Phys. Rev. Research {\bf 4}, 023207 (2022). 


\bibitem{agarwal2} D. Bhatti, R. Schneider, S. Oppel and J. von Zanthier , Directional Dicke Subradiance with Nonclassical and Classical Light Sources, Phys. Rev. Lett. {\bf 120}, 1136 (2018).

\bibitem{agarwal3} J. Xu, S. Chang, Y. Tang, S. Zhu, G. S. Agarwal, Hyperradiance accompanied by nonclassicality, Phys. Rev. A {\bf 96}, 013839 (2017). 

\bibitem{agarwal4} M. Pleinert, J. von Zanthier, and G. S. Agarwal, Hyperradiance from collective behavior of coherently driven atoms, Optica {\bf 4}, 779 (2017). 

\bibitem{kimble} A. Asenjo-Garcia, M. Moreno-Cardoner, A. Albrecht, H.J. Kimble, and D. E. Chang, Exponential Improvement in Photon Storage Fidelities Using Subradiance and “Selective Radiance” in Atomic Arrays, Phys. Rev. X {\bf 7}, 031024 (2017).

\bibitem{browaeys4} G. Ferioli, A. Glicenstein, I. Ferrier-Barbut and
A. Browaeys, A non-equilibrium superradiant phase
transition in free space, Nature Physics {\bf 19}, 1345 (2023).

\bibitem{yan} Z. Yan, J. Ho,  Y. Lu, S. J. Masson,
A. Asenjo-Garcia, and D. M. Stamper-Kurn, Superradiant and Subradiant Cavity Scattering by Atom Arrays, Phys. Rev. Lett. {\bf 131}, 253603 (2023). 

\bibitem{mandel} L. Mandel and E. Wolf, Optical Coherence and Quantum Optics (Cambridge University Press, 1995). 

\bibitem{ising1} F. Barahona, On the Computational Complexity of Ising Spin-Glass Models, J. Phys. A: Math. Gen. {\bf 15}, 3241 (1982). 

\bibitem{ising2} D. L. Stein and C. M. Newman, Spin Glasses: Old and New Complexity, arXiv:1205.3432 [cond-mat.stat-mech] (2012). 


\end{references}
\end{document}